\documentclass[aps,pra,twocolumn,a4paper,showpacs,superscriptaddress,floatfix,10pt]{revtex4}

\usepackage{graphicx}
\usepackage{amsmath}
\usepackage{epsfig}
\usepackage{helvet}
\usepackage{amssymb}
\usepackage{epstopdf}
\usepackage{color}
\usepackage{bbold}
\usepackage{cancel}

\begin{document}

\title{Interaction stabilized steady states in the driven $O(N)$ model}

\author{Anushya Chandran}
\affiliation{Perimeter Institute for Theoretical Physics, Waterloo, Ontario N2L 2Y5, Canada}   \email{achandran@perimeterinstitute.ca}
\author{S. L. Sondhi}
\affiliation{Department of Physics, Princeton University, Princeton, NJ 08544}
\affiliation{Max-Planck-Institut f\"ur Physik komplexer Systeme, Dresden, Germany}

\date{\today}
\begin{abstract}
We study periodically driven bosonic scalar field theories in the infinite $N$ limit.
It is well-known that the free theory can undergo parametric resonance under monochromatic modulation of the mass term and thereby absorb energy indefinitely.
Interactions in the infinite $N$ limit terminate this increase for any choice of the UV cutoff and driving frequency.
The steady state has non-trivial correlations and is synchronized with the drive. 
The $O(N)$ model at infinite $N$ provides the first example of a clean interacting quantum system that does not heat to infinite temperature at any drive frequency.
\end{abstract}

\maketitle

\section{Introduction}

Experiments in cold atomic systems have generated much interest in the dynamics of periodically driven many-body Hamiltonians \cite{Bloch:2008ly, Polkovnikov:2011ys}.
Energy is not conserved in such systems; instead by the Bloch-Floquet theorem, the eigenstates have the form: 
\begin{align}
\label{Eq:FloqEig}
|\psi(t)\rangle = e^{-i \epsilon t} |\phi(t)\rangle
\end{align}
where $\epsilon$ is a quasi-energy defined modulo the fundamental drive frequency and $|\phi(t)\rangle$ has the same periodicity as the drive \cite{Shirley:1965ff}.
There are two intimately related questions in such a system: i) is there a late time steady state with non-trivial correlations and finite energy density, and ii) is such a late-time steady state synchronized with the drive \footnote{By synchrony with the drive, we mean that observables are periodic in time with the same period as the drive.}? Already for a single two-level system, the answers to these two questions are non-trivial as the system can coherently Rabi flip-flop at a different frequency from that of the drive.

At the next level of complexity are integrable many-body systems such as periodically driven non-interacting fermions.
The dynamics is governed by an effective quadratic Floquet Hamiltonian; thus the stationary state coincides with an appropriate periodic generalized Gibbs ensemble \cite{Russomanno:2012qe, Lazarides:2014kq}.
There has been much theoretical progress classifying the topological structure of Floquet bands \cite{Rudner:2013mz,Nathan:2015gd,Roy:pi} and experimental progress studying such states in cold atomic systems \cite{Jotzu:2014bh} and topological insulators \cite{Wang:2013fv, Onishi:2014qf}.

The nature of the steady state in driven interacting systems is less clear.
Standard linear response theory suggests that any finite frequency drive heats the system to infinite temperature. 
That is, the local reduced density matrix approaches the identity.
Refs.~\cite{Ponte:2015zr,DAlessio:2014ty,Lazarides:2014yg} argue for this scenario in generic ergodic systems with locally bounded Hilbert spaces.
These findings are in contradiction with Refs.~\cite{Prosen:1998hc, Prosen:1999rc, DAlessio:2013rm} that claim that certain spin models do not heat to infinite temperature when the drive frequency is above a finite threshold.
For strongly disordered spin systems whose time independent Hamiltonian is many-body localized, several recent studies find the same threshold behavior \cite{Ponte:2015zp, Lazarides:2014rq, Abanin:2014fu}.
For systems with locally unbounded Hilbert spaces, even less is known.
One recent study of the many-body Kapitza pendulum finds threshold behavior~\cite{Citro:2015aa}; we compare their results to ours later.

In this article, we approach this problem using the large $N$ expansion for interacting bosonic systems.
The $O(N)$ model at infinite $N$ is a canonical model for symmetry-breaking in statistical mechanics \cite{Moshe:2003aa}.
Its equilibrium properties are exactly soluble and capture the correct topology of phase diagrams in various dimensions.
It is also a canonical model for the unitary dynamics of interacting theories and a workhorse of many fields including cosmology and condensed matter \cite{Bray:1994aa,Boyanovsky:1996kx,Cooper:1997mi,Boyanovsky:2000aa,Sotiriadis:2010ys,Das:2012kx,Sciolla:2013aa,Chandran:2013aa}.
The Floquet dynamics in the infinite $N$ limit is the focus of this work.
We comment on $1/N$ corrections towards the end, but reserve a full-blown treatment for the future.

We begin in Sec.~\ref{Sec:Gaussian} by reviewing the response of the Gaussian model under periodic driving of the bare mass in the paramagnetic phase, previously studied in Refs.~ \cite{Ferrari:1998aa, Weigert:2002aa}.
In this theory, the energy density grows exponentially in time via the parametric resonance of selected momentum modes \cite{Landau:1980xq}.
Incorporating the effect of interactions at $N=\infty$ removes this divergence for any fixed cutoff and drive frequency (Sec.~\ref{Sec:PM}).
The long-time steady state synchronizes with the drive, but depends on initial conditions because of the integrability of the theory.
In Sec.~\ref{Sec:FM}, we turn to the symmetry-broken phase. 
Again, interactions prevent indefinite heating, although the time averaged magnetization decays to zero.
Oddly, the late time magnetization can oscillate at half the frequency of the drive; we argue that this is an artifact of the infinite $N$ limit.
The $O(N)$ model at infinite $N$ is the first example of a many-body system that fails to heat to infinite temperature for any periodic drive. 

\section{Gaussian model}
\label{Sec:Gaussian}
\begin{figure}[tbp]
\begin{center}
\includegraphics[width=0.9\columnwidth]{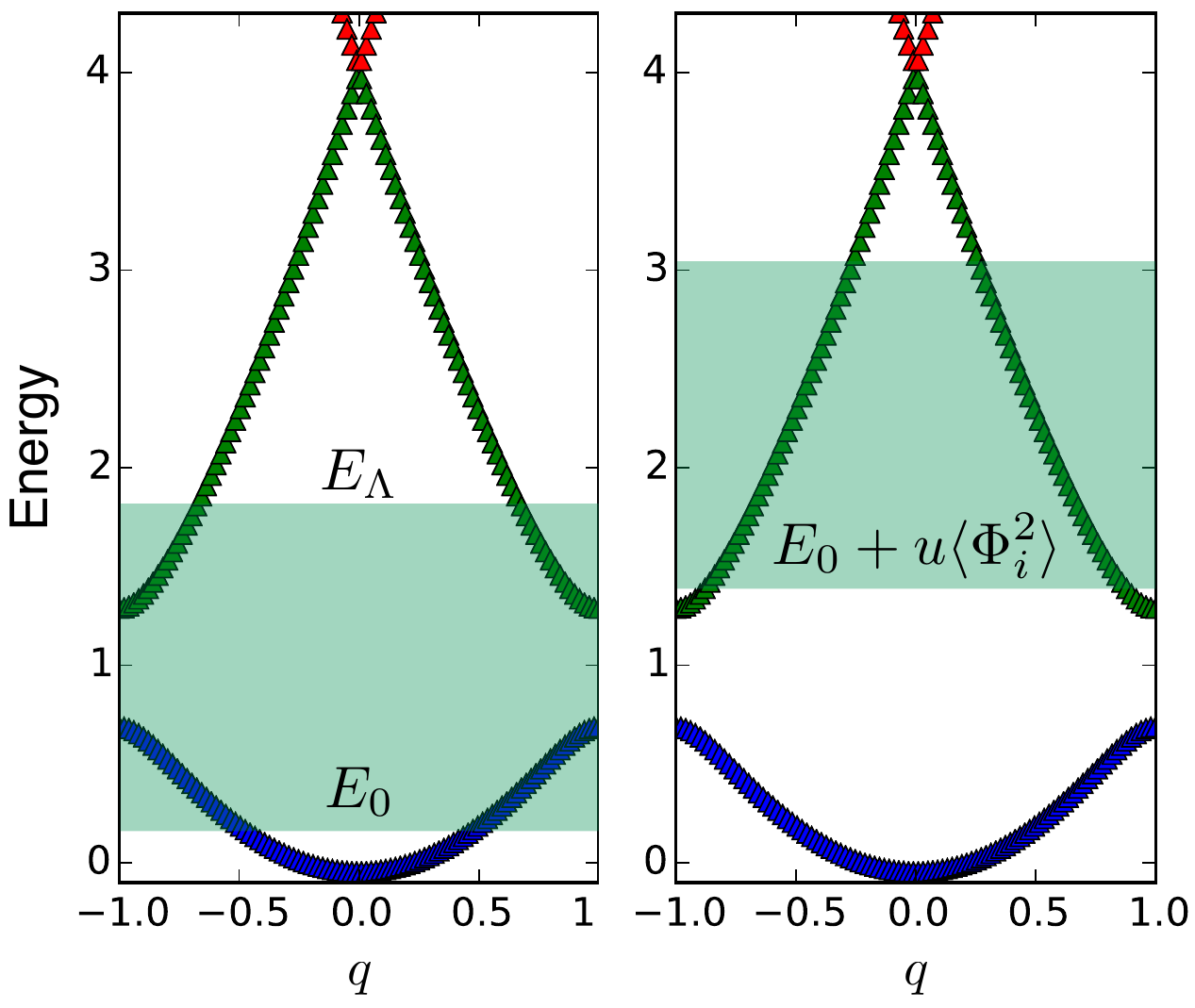}
\caption{Left: the energy spectrum of a $1d$ particle in a cosine potential vs quasi-momentum $q$. The shaded region is the mode range. If the mode range includes band-gap(s), the Gaussian model is unstable. Right: the effective Mathieu spectrum in the driven $O(N)$ model at late times. The mode range always lies within a band.}
\label{Fig:Fig1MatSpec}
\end{center}
\end{figure}

Consider the free $O(N)$ model in $d$ dimensions with a sinusoidally varying bare mass. 
The Hamiltonian is:
\begin{align}
H_0(t) &= \int^\Lambda \frac{d^d k}{(2\pi)^d} \left(\frac{|\Pi_i(\vec{k})|^2}{2} +  (|\vec{k}|^2 + r(t) ) \frac{|\Phi_i(\vec{k})|^2}{2}\right) \nonumber\\
r(t) &= r_0 - r_1 \cos(\gamma t)
\end{align} 
where $\Phi_i(\vec{k})$ and $\Pi_j(\vec{k})$ satisfy the canonical commutation relations:
 \begin{align}
 [\Phi_i(\vec{k}), \Pi_j(\vec{k'})] = i (2\pi)^d \delta(\vec{k}-\vec{k'}) \delta_{ij}.
 \end{align}
The component index $i$ runs from $1$ to $N$, the ultraviolet cut is given by $\Lambda$ in momentum space and the bare mass is $r(t)$.
We suppress the label $i$ as the initial states are $O(N)$-symmetric.
In equilibrium in the absence of drive ($r_1=0$), the model is paramagnetic for $r_0 \geq 0$ at all temperatures and ill-defined for $r_0 <0$.

In the bulk of the article, we analyze the response under a monochromatic drive at a single frequency $\gamma$.
As the Gaussian theory is a linear system, the response to a polychromatic drive follows by superposition.
The interacting $O(N)$ model is however not a linear system.
We address generic periodic drives in this model at the end of Sec.~\ref{Sec:PM}.

In the Gaussian model, the equations of motion for $\Phi(\vec{k},t)$, $\Pi(\vec{k},t)$ are linear.
We expand these operators in a fixed basis of creation and annihilation operators, e.g. $\Phi(\vec{k}, t) = f_{\vec{k}}(t) a_{\vec{k}} + f^*_{-\vec{k}}(t) a^\dagger_{-\vec{k}}$.
This defines the complex mode functions $f_{\vec{k}}(t)$.
The commutation relation between $\Phi(\vec{k},t)$ and $\Pi(\vec{k},t)$ imposes the following constraint:
\begin{align}
\label{Eq:Wronskian}
\textrm{Im}[f_{\vec{k}}(t) \dot{f}^*_{\vec{k}}(t)] = 1/2 \quad\forall \vec{k}
\end{align}
The mode functions satisfy the equations of motion:
\begin{align}
\label{Eq:OrigMode}
\left( \frac{d^2}{dt^2} +  |\vec{k}|^2 + r_0 - r_1 \cos(\gamma t) \right) f_{\vec{k}} = 0
\end{align}
At each momentum $\vec{k}$, this is the well-known equation of motion of a parametrically driven two-dimensional harmonic oscillator (as $f_{\vec{k}}$ is complex).
The transformation, $t \rightarrow 2 t/\gamma$, $f_{\vec{k}} \rightarrow \sqrt{2/\gamma} f_{\vec{k}}$, makes time and the mode function dimensionless and brings Eq.~\eqref{Eq:OrigMode} to the canonical form:
\begin{align}
\label{Eq:Mathieu}
\left( \frac{d^2}{dt^2} +  E_k - 2g \cos(2t) \right) &f_{\vec{k}} = 0\\
E_k = \frac{4( |\vec{k}|^2 + r_0 )}{\gamma^2},\quad &g = \frac{2 r_1}{\gamma^2} \label{Eq:Ekq}
\end{align}
This differential equation is the Mathieu equation \footnote{$\textrm{Im}[f_{\vec{k}}(t) \dot{f}^*_{\vec{k}}(t)] $ is an invariant of the Mathieu equation, so that the commutation relations hold for all times if they hold at the initial time.}.

The Mathieu equation is a familiar beast in band theory; it is the Schr{\"o}dinger equation of a one-dimensional particle in a cosine potential. 
This identification provides a dictionary between the mode functions $f_{\vec{k}}(t)$ and the Bloch wavefunctions of the Schr{\"o}dinger equation:
\begin{align}
t &\rightarrow \textrm{Spatial coordinate} \nonumber\\
f_{\vec{k}}(t) &\rightarrow \textrm{Wavefunction} \nonumber\\
E_k &\rightarrow \textrm{Energy of the wavefunction} \label{Eq:MapMathieu}
\end{align}
A number of properties of the Mathieu spectrum (Fig.~\ref{Fig:Fig1MatSpec}) follow directly from this mapping:
\begin{enumerate}
\item By the Bloch theorem, the spectrum of the particle is labelled by a quasi-momentum $q \in (-1, 1]$ and a band index $m=1,2\ldots$ with eigenfunctions: $\psi(x) = e^{-iqx} \phi^m(x) $ where $\phi^m(x) = \phi^m(x+\pi)$.
\item When the amplitude of the drive $g$ is zero, the spectrum is degenerate at the centre and the edges of the Brillioun zone. Any $g\neq 0$ opens a gap at these degenerate points; the $m$th gap is approximately $g^{m}/((m/2)!)^2$ for large $m$. The gaps decrease faster than exponentially in $m$.
\item The bandwidth of the $m$th band is approximately $2m+1$ at large $m$.
\end{enumerate}

Using the mapping in Eq.~\eqref{Eq:MapMathieu}, the $k$-modes in the Gaussian theory sample energies from $E_0 = 4r_0/\gamma^2$ to $E_\Lambda = 4(r_0+\Lambda^2)/\gamma^2$ in the Mathieu spectrum (shaded region in Fig.~\ref{Fig:Fig1MatSpec}).
We call the range of energies between $E_0$ and $E_\Lambda$ as the `mode range'.
If the mode range lies within a Mathieu band, then by (1) above, each mode function is a superposition of the two solutions at $\pm q_k$ and is oscillatory in time:
\begin{align}
\label{Eq:GenMatEq}
f_{\vec{k}}(t) = \alpha_k e^{-i q_k t} \phi^m_{\vec{k}}(t) + \beta_k e^{i q_k t} \phi^m_{\vec{k}}(t)^* 
\end{align}
where $\alpha_k, \beta_k$ are complex numbers determined by the initial conditions and Eq.~\eqref{Eq:Wronskian}.
The energy density and other spatially local observables involve integrals over the mode functions in $k$-space.
As the magnitude of each mode function is bounded in Eq.~\eqref{Eq:GenMatEq}, all such observables remain bounded as $t\rightarrow \infty$.
Further, it is straightforward to show that local observables oscillate in synchrony with the drive as $t\rightarrow \infty$.
For example, $\langle \Phi_i^2(t) \rangle$ is given by:
\begin{align*}
\langle \Phi_i^2(t) \rangle = \int^\Lambda \frac{d^dk}{(2\pi)^d} \, \delta_k |\phi^m_{\vec{k}}(t)|^2 + 2 \textrm{Re}[ e^{-2i q_k t} \chi^m_{\vec{k}}(t)]
\end{align*}
where $\delta_k$ and $\chi^m_{\vec{k}}(t)$ are related to the parameters in Eq.~\eqref{Eq:GenMatEq}.
The second term decays as $1/t^{d/2}$ as $t\rightarrow \infty$ while the first term has the same period $T=\pi$ as the drive.
Thus, $\langle \Phi_i^2(t) \rangle = \langle \Phi_i^2(t+\pi) \rangle$ as $t\rightarrow \infty$.

\begin{figure}[tbp]
\begin{center}
\includegraphics[width=0.7\columnwidth]{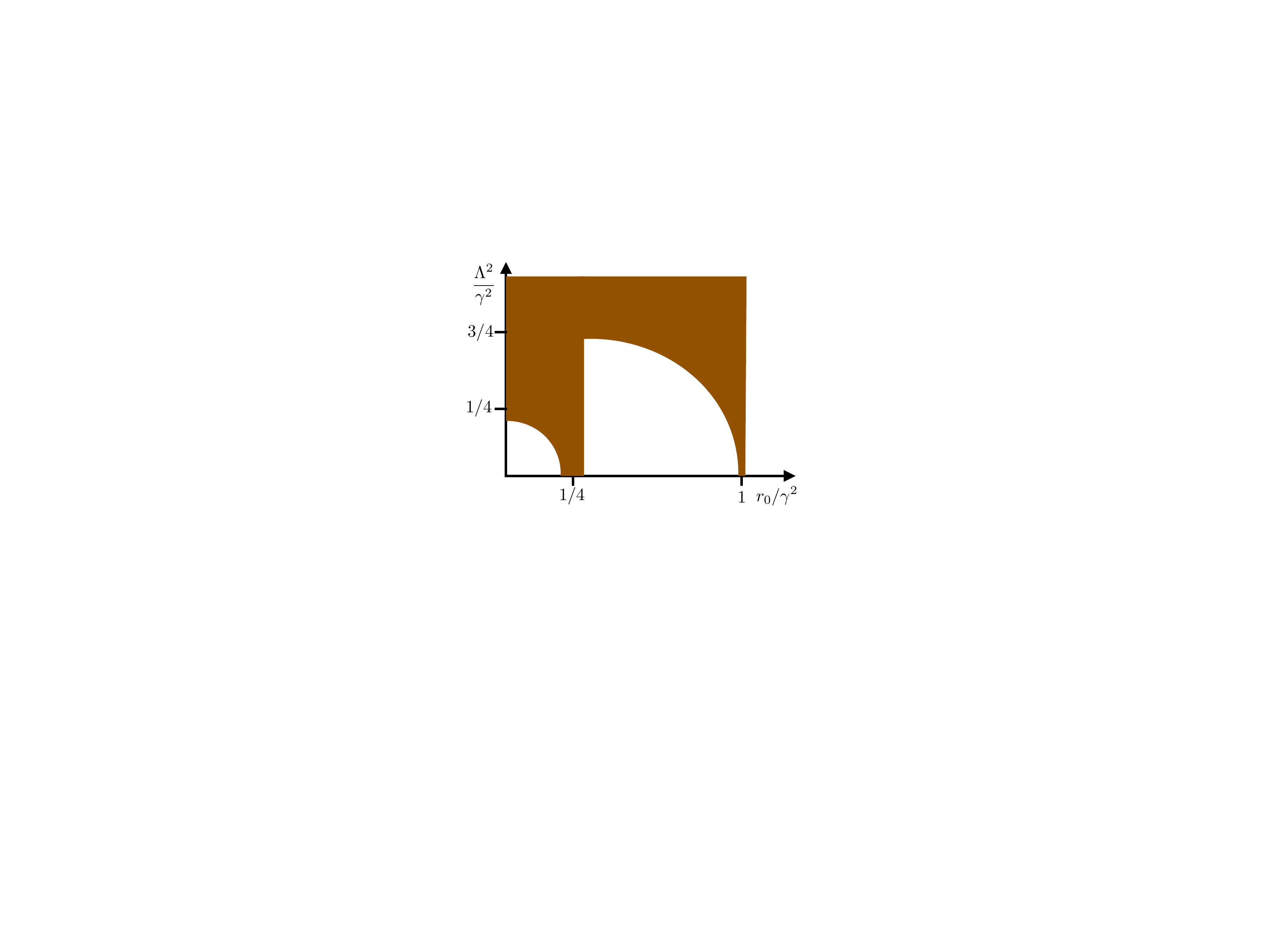}
\caption{The dynamical phase diagram of the driven Gaussian theory at fixed $g/\gamma^2$. Unshaded/shaded regions are stable/unstable. The band gaps in the Mathieu spectrum determine the phase boundaries. }
\label{Fig:FigPhaseDiaGM}
\end{center}
\end{figure}

The phase diagram is sketched in Fig.~\ref{Fig:FigPhaseDiaGM}.
The simplest stable phase lies at small $\Lambda^2/\gamma^2, r_0/\gamma^2$; here the drive frequency $\gamma$ is much greater than any other energy scale in the system.
However, the additional stable phases are non-trivial consequences of the band structure of the Mathieu equation.

When the mode range intersects the Mathieu band-gaps, the corresponding mode functions increase exponentially in time, exhibiting parametric resonance.
Consequently, the energy density (and other local observables) also grow exponentially in time and the system heats indefinitely. 
The heating time-scale is given by the inverse of the largest band-gap intersecting with the mode range.
It is important to note that at any fixed bare mass and drive parameters, the theory is always unstable for sufficiently large cutoff $\Lambda$ as the mode range increases with $\Lambda$.

It is sometimes useful to think of the evolution over a period $T=2\pi/\gamma$ as being generated by an effective Floquet Hamiltonian $H_F$:
\begin{align}
U(T) = e^{-i H_{F} T}
\end{align}
where $U(T)$ is the evolution operator for a period \cite{Shirley:1965ff, Sambe:1973tk}.
As the theory is quadratic, $H_F$ can be chosen to be quadratic in the field operators.
In the stable regime, the mode spectrum of $H_F$ is non-negative and the eigenmodes are normalizable.
By expanding any initial state in this eigenbasis, it is easily seen that the late time response is stable and periodic.
In the unstable regime, on the other hand, the mode spectrum includes negative energies and corresponding unnormalizable eigenmodes.
This is what allows the system to absorb energy indefinitely.
For more details, see Ref.~\cite{Weigert:2002aa}.

We end with three comments.
First, the physics discussed above applies to any spatial dimension $d\geq 1$.
Second, in the phase diagram in Fig.~\ref{Fig:FigPhaseDiaGM}, the stable region persists to some $r_0/\gamma^2<0$.
Thus, the driven Gaussian theory can be stable even when the equilibrium theory is not.
Finally, ``energy density" when unqualified refers to either the instantaneous energy density or the energy density with respect to the time-averaged Hamiltonian. Both diverge when the system heats up to infinite temperature.

\section{Driven Paramagnet}
\label{Sec:PM}
We now turn to the interacting $O(N)$ model with a sinusoidally varying bare mass. The Hamiltonian reads:
\begin{align}
H(t) =  H_0(t) + \frac{\lambda}{4N} \sum_{i=1}^N \int d^dx  \,(\Phi_i(\vec{x}))^4
\end{align}
At infinite $N$, $( \Phi_i(\vec{x}) \Phi_i(\vec{x}) )/{N}$ acts like a classical time-dependent field and can be replaced by its expectation value.
In the absence of the drive in equilibrium ($r_1=0$), the model is paramagnetic for all $r_0 > r_c$ and spontaneously breaks the $O(N)$ symmetry for $r_0< r_c$.
The value of $r_c$ is determined by $d$: in $d=1$, $r_c = -\infty$, while in $d\geq 2$, $r_c$ is negative and finite.
Further, the symmetry-broken phase extends to finite temperatures for $d\geq 3$.
In this section, we focus on the coherently driven paramagnet.

Expanding in a fixed basis of creation/annihilation operators as before, we obtain the equations of motion:
\begin{align}
\left( \frac{d^2}{dt^2} +  |\vec{k}|^2 + r(t) +  \lambda\int^\Lambda \frac{d^dk}{(2\pi)^d} |f_{\vec{k}}(t)|^2 \right) f_{\vec{k}}(t) = 0 \label{Eq:LargeNOrigEom}
\end{align}
assuming that $f_{\vec{k}}(t)$ is the same for every component and using the relation (no summation on $i$):
\begin{align}
\langle \Phi_i^2(t)  \rangle = \int \frac{d^dk}{(2\pi)^d} \, |f_{\vec{k}}(t)|^2 \label{Eq:Phi2Def}
 \end{align}
As the system is spatially homogenous, $\langle \Phi^2_i\rangle$ only depends on $t$. For more details, see Ref.~\cite{Chandran:2013aa}.
As a consequence of the quartic term, Eq.~\eqref{Eq:LargeNOrigEom} is non-linear in the mode-functions. 
Re-writing in dimensionless units:
\begin{align}
\label{Eq:LargeNcanonicalPM}
\left( \frac{d^2}{dt^2} +  E_k - 2g \cos(2t) + u \int^\Lambda \frac{d^dk}{(2\pi)^d} |f_{\vec{k}}(t)|^2  \right) f_{\vec{k}}(t) = 0 
\end{align}
where $u \equiv 8\lambda/\gamma^3$ and $E_k, g$ are defined in Eq.~\eqref{Eq:Ekq}.

Observe that $u=0$ corresponds to the driven Gaussian model.
When the Gaussian model exhibits stable behavior, it is clear that a small $u$ merely dresses the steady state. 
The main result is that the stability persists at all parameters (even when the Gaussian model is unstable).
The energy density \emph{always} plateaus to a finite value at late times and the wavefunction has non-trivial correlations that can be described within an effective Gaussian model.
Below, we present the intuition underlying this stability and numerical results that support this claim.
We then construct explicit Floquet solutions to Eq.~\eqref{Eq:LargeNcanonicalPM} at low drive frequency within the WKB approximation.

The quartic term acts as a self-consistent correction to the energy $E_k$ in Eq.~\eqref{Eq:LargeNcanonicalPM}.
Define the instantaneous energy: 
\begin{align}
\label{Eq:EffEEffq}
E_k(t) = E_k + u \langle \Phi_i^2(t) \rangle,
\end{align}
This identifies an instantaneous mode range.
Suppose that at $t=0$, the mode range includes band-gaps in the Mathieu spectrum (Fig.~\ref{Fig:Fig1MatSpec}).
Then, the associated mode functions grow exponentially in time, and by Eq.~\eqref{Eq:Phi2Def}, so does $\langle \Phi_i^2(t)\rangle$.
Finally, this implies the mode range itself drifts up with time.
As the bandwidth of each Mathieu band is proportional to its index, at large enough $E_0(t)$, the mode range lies within a single band (Fig.~\ref{Fig:Fig1MatSpec}).
We expect that all the mode functions then become oscillatory, the time averaged value of $\langle \Phi_i^2(t) \rangle$ plateaus, and the system stops heating.
Further, just as in the Gaussian model, we expect that all observables synchronize with the drive.

A more refined version of the above argument corrects the drive at late times.
At late times, $\langle \Phi_i^2(t) \rangle$ is in synchrony with the drive:
\begin{align}
\label{Eq:FSpecPhi2}
\langle \Phi_i^2(t) \rangle \sim \sum_{\omega=0}^\infty F[\langle \Phi_i^2 \rangle ](\omega) e^{2 i \omega t}
\end{align}
where $F[\langle \Phi_i^2 \rangle ](\omega)$ denotes the Fourier amplitude at frequency $\omega$.
The mode functions thus satisfy a generalized Mathieu equation with parameters:
\begin{align}
\label{Eq:PMEffParams}
\bar{E}_k= E_k + u F[\langle \Phi_i^2 \rangle](0), \quad \bar{g} = g - u F[\langle \Phi_i^2\rangle](2)
\end{align}
The higher harmonics in $\langle \Phi_i^2(t) \rangle$ can be ignored as their amplitudes are small as compared to $\bar{g}$.
We expect that the mode range lies within a single band of the Mathieu spectrum with the parameters in Eq.~\eqref{Eq:PMEffParams}.

The arguments above, while appealing, are not decisive.
At short times, they assume that Eq.~\eqref{Eq:LargeNcanonicalPM} can be treated as an effective Mathieu equation even though $\langle \Phi_i^2(t)\rangle$ increases exponentially in time. 
At long times, they ignore the higher harmonics in the effective drive.
We therefore turn to numerical simulations to confirm this picture.

\begin{figure}[tbp]
\begin{center}
\includegraphics{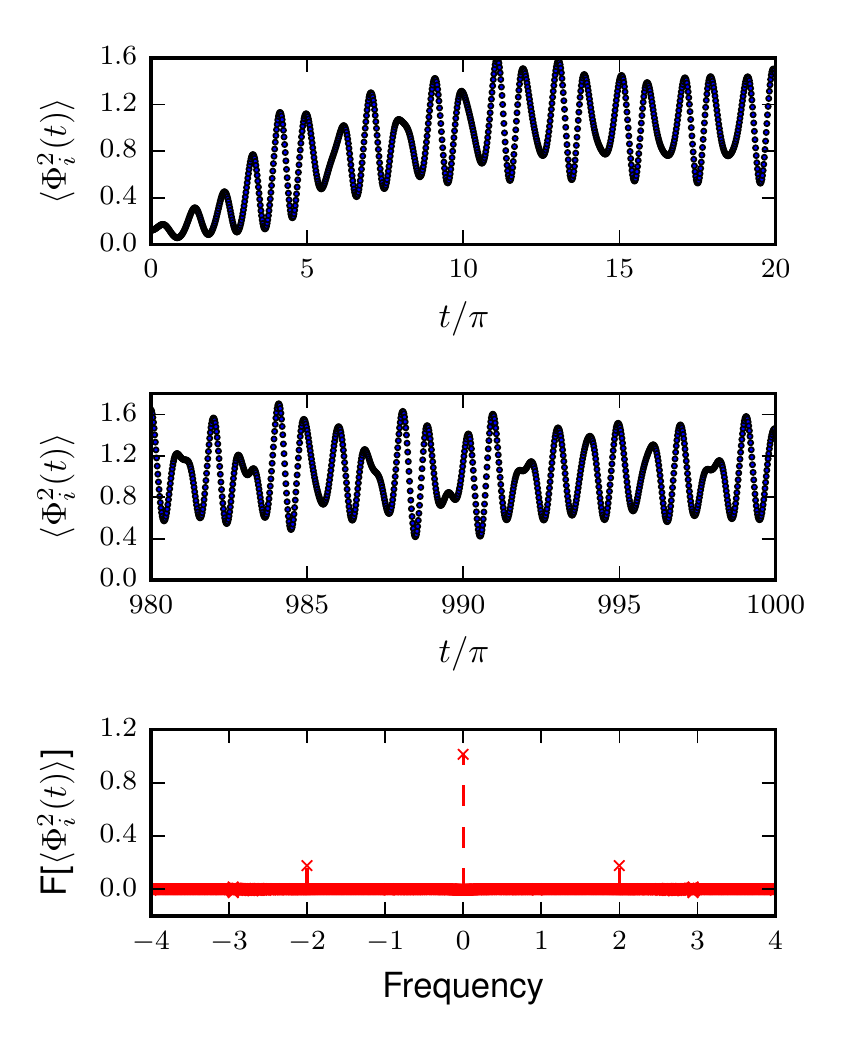}
\caption{Top and middle panels: $\langle \Phi_i^2(t)\rangle$ vs $t/\pi$ at early and late times when the paramagnetic ground state is driven starting at $t=0$. Bottom panel: Fourier spectrum of $\langle \Phi_i^2(t)\rangle$ over a thousand periods of the drive. Parameters: $d=1$, $r_0=0.1$, $u=1$, $\Lambda=1$, $g=0.3$ and system size $L=100$. The energy density shows similar behavior.}
\label{Fig:Phi2ParaON}
\end{center}
\end{figure}

\paragraph{Numerics---}
Consider the $d=1$ $O(N)$ model in its paramagnetic ground state at $t=0$.
At time $t=0$, we turn on the drive.
With the parameters chosen in Fig.~\ref{Fig:Phi2ParaON}, the mode range at $t=0$ includes the first band-gap.
The modes near the center of the band gap set the initial time-scale of exponential growth of $\langle \Phi_i^2(t)\rangle$ to be approximately $2/g$.
The width of the mode range is $1$, while the bandwidth of the second Mathieu band is approximately $3$; thus we expect $\langle \Phi_i^2(t)\rangle$ to stop growing when $\bar{E}_0$ is approximately the energy at the bottom of the second band of the effective spectrum for the parameters in Eq.~\eqref{Eq:PMEffParams}.

In Fig.~\ref{Fig:Phi2ParaON}, we plot $\langle \Phi_i^2(t)\rangle$ vs the dimensionless time $t$ at early (top) and late (middle) times.  
With $g=0.3$, the initial time-scale of exponential growth is approximately $7$, in good agreement with the top panel in Fig.~\ref{Fig:Phi2ParaON}.
The growth rate is seen to decrease with time, as predicted by the effective Mathieu picture. 
At $t/\pi\approx 10$, $\langle \Phi_i^2(t)\rangle$ saturates.
This is evidence for stability; indeed, $\langle \Phi_i^2(t)\rangle$ oscillates about a mean value for the next thousand periods (middle panel).
To connect to the effective Mathieu spectrum, we plot the Fourier spectrum (bottom panel).
We see that i) $\langle \Phi_i^2(t)\rangle$ is synchronized with the drive (Eq.~\eqref{Eq:FSpecPhi2}), ii) the harmonics at frequencies above two are suppressed, and iii) the value of $\bar{E}_0$ extracted from the figure coincides with the value of the lowest energy in the second band of the effective Mathieu spectrum.
The energy density (not shown) is also finite and synchronized with the drive as $t\rightarrow \infty$.
Thus, the numerical simulations confirm the effective Mathieu picture and provide evidence for a steady state stabilized by interactions in the driven $O(N)$ model. 

\paragraph{Approximate Floquet solutions---}We construct explicit normalizable solutions of Eq.~\eqref{Eq:LargeNcanonicalPM} as the drive frequency $\gamma$ approaches zero within the WKB approximation \cite{Bender:1999kx}.
Physically, the system is most susceptible to indefinite heating at low drive frequency; the construction of normalizable Floquet states in this limit is strong evidence that the model exhibits stable behavior at any drive parameters.

Following the structure in Ref.~\cite{Bender:1999kx}, we re-arrange Eq.~\eqref{Eq:LargeNcanonicalPM} as:
\begin{align}
\label{Eq:WKB}
\left( \frac{\gamma^3}{4} \frac{d^2}{dt^2} +  \gamma(|\vec{k}|^2  + r_0 - r_1 \cos(2t)) + 2\lambda \langle \Phi_i^2(t) \rangle  \right) f_{\vec{k}}(t) = 0 
\end{align}
Note that we are still working in dimensionless units.
To leading order in $\gamma$ in the WKB series, the mode functions are independent of $\vec{k}$ and are given by: 
\begin{align*}
f_{\vec{k}}(t) \sim \frac{1}{\sqrt{2 \sqrt{8 \lambda \langle \Phi_i^2(t) \rangle }}} \exp\left[ \pm i \int^t \sqrt{8 \lambda \langle \Phi_i^2(t') \rangle} dt'\right]
\end{align*}
where $\langle \Phi_i^2 (t) \rangle$ is determined by the self-consistency condition in Eq.~\eqref{Eq:Phi2Def}:
\begin{align}
\langle \Phi_i^2(t) \rangle \sim K \frac{\Lambda^{2d/3}}{\lambda^{1/3}}
\end{align}
$K$ is a dimension dependent constant.
To next order in $\gamma$, the mode functions depend on $\vec{k}$ and $\langle \Phi_i^2(t)\rangle$ is corrected by a $\gamma \cos(2t)$ term.
This expansion is systematic; we may go to as high an order in $\gamma$ as we desire and construct asymptotically accurate oscillatory solutions to Eq.~\eqref{Eq:LargeNcanonicalPM}.
 
\paragraph{Comments---} 
We end with two comments. 
First, the effective Mathieu picture can be generalized to the case of a polychromatic periodic drive with fundamental frequency $\gamma$.
This is because the higher harmonics at frequencies $\pm m \gamma$, $m\geq 2$ do not change the qualitative picture of bands in Fig.~\ref{Fig:Fig1MatSpec}.
As in the monochromatic case, the mode range drifts up in time until it lies within a single band of the effective Mathieu equation.
Thus, the driven $O(N)$ model reaches an interaction stabilized synchronized steady state for any periodic drive.

Second, the late time steady state is not described by a single temperature.
Instead, it is described by an emergent periodic generalized Gibbs ensemble (PGGE) associated with the effective Gaussian model  \cite{Chandran:2013aa}.
The conserved quantities that feature in the PGGE are the mode occupations of the effective Gaussian model with parameters in Eq.~\eqref{Eq:PMEffParams}.
Let $n_{\vec{k}}(t)$ denote the mode occupation of the oscillator at momentum $\vec{k}$.
At late times, $n_{\vec{k}}(t)$ commutes with the unitary evolution operator $U(t)$ and is periodic with the same period as the drive.
As these quantities are conserved only at late times, the PGGE is emergent. 

\section{Driven Ferromagnet}
\label{Sec:FM}

Like the driven paramagnet, the driven ferromagnet does not heat to infinite temperature at any drive parameters.
Rather, it reaches a stable paramagnetic steady state in any $d\geq 2$.
Unlike the driven paramagnet however, the time period of the observables at late-time can be double that of the drive.
We believe that this lack of synchrony is a consequence of the infinite $N$ limit, as we discuss below.

Let us be more precise.
For $r_0< r_c$ in $d\geq 2$, the $O(N)$ symmetry is spontaneously broken in equilibrium.
Let the symmetry be broken along the $1$ direction in order parameter space so that $\langle \Phi_1(\vec{x}) \rangle$ is the non-zero uniform magnetization. 
The excitations along the $1$ direction are massive. The remaining $(N-1)$ directions in order parameter space are soft and support Goldstone modes.
The system is in the ground state at $t=0$ when the mass drive is switched on.

At infinite $N$, the equations of motion involve two classical fields: $\langle \Phi_1(t) \rangle$ and $\sum_{i=2}^N \langle \Phi_i^2(t)\rangle/N$.
The new classical field defines the magnetization $M(t) \equiv \langle \Phi_1(t) \rangle/\sqrt{N}$.
Defining mode functions for components $2,\ldots N$ as before and going to dimensionless units ($t \rightarrow 2 t/\gamma$, $f_{\vec{k}} \rightarrow \sqrt{2/\gamma} f_{\vec{k}}$, $M(t) \rightarrow \sqrt{2/\gamma} M(t)$), the equations of motion are:
\begin{align}
\left( \frac{d^2}{dt^2} +  E_k - 2g \cos(2t) + r_f(t) \right) &f_{\vec{k}}(t) = 0 \\
\left( \frac{d^2}{dt^2} +  E_0 - 2g \cos(2t) + r_f(t) \right) &M(t) = 0 \label{Eq:Meom}\\
r_f(t) \equiv u \int^\Lambda \frac{d^dk}{(2\pi)^d} |f_{\vec{k}}(t)|^2 + &u M^2(t) \label{Eq:rfdef}
\end{align}
where $E_k, g$ are defined in Eq.~\eqref{Eq:Ekq} and $u \equiv 8\lambda/\gamma^3$ as in the previous section.
The feedback term to the bare mass is denoted by $r_f(t)$ and involves an extra classical field as compared to Eq.~\eqref{Eq:LargeNcanonicalPM}.

The intuitive argument for stability at late times is analogous to the one in Sec.~\ref{Sec:PM} if we define the instantaneous energy as $E_k(t) \equiv E_k + r_f(t)$.
At $t=0$, the mode range extends from $E_0(0)=0$ to $E_\Lambda(0)$.
$E_0(0)$ is exactly zero as the Goldstone modes are massless.
If the mode range lies within the lowest band, then the solution is stable as the mode functions are oscillatory (Sec.~\ref{Sec:Gaussian}).
If however the mode range includes band-gaps, then $r_f(t)$ increases in time until the mode range fits in a single band of an effective Mathieu spectrum with parameters:
\begin{align}
\bar{E}_k = E_k +  F[  r_f ](0) \quad \bar{g} = g -  F[ r_f ](2)
\label{Eq:EffMatEqFM}
\end{align}
Again, $F[  r_f ](\omega)$ denotes the Fourier amplitude at frequency $\omega$. 

The first consequence of the above picture is that $\bar{E}_0$ exactly coincides with the bottom of an effective Mathieu band in the driven ferromagnet.
In contrast, in the driven paramagnet, $\bar{E}_0$ could lie anywhere within an effective band as long as the mode range fits in the band.
A simple example is an initial state with $E_0(0) =2$ and $E_\Lambda(0)=3$ for the parameters chosen in Fig.~\ref{Fig:Phi2ParaON}.
As the example suggests, the difference stems from the tunability of $E_0(0)$ in the paramagnetic case.

A second consequence is that the time period of late-time observables can be twice the period of the drive.
This is because $M(t)$ satisfies an effective Mathieu equation with parameters $\bar{E}_0$ and $\bar{g}$ at late times and $\bar{E}_0$ coincides with the bottom of an effective band.
If the bottom of the band corresponds to quasi-momentum $q=0$ (mod $2$) in Fig.~\ref{Fig:Fig1MatSpec}, then observables have the same period as the drive at late times.
If instead the quasi-momentum is $1$ (mod $2$), the period is twice that of the drive.
As the oscillations themselves are likely an artifact of the infinite $N$ approach, we do not expect this behavior at any finite $N$ (see Sec.~\ref{Sec:Discussion}).

Finally, the time average of $M(t)$ is zero at late times, as it is the solution of a Mathieu equation.
Thus the steady state is paramagnetic. 
Interestingly, this is true in any dimension $d\geq 2$, even when there is a finite temperature ordered phase in equilibrium.

\begin{figure}[tb]
\begin{center}
\includegraphics{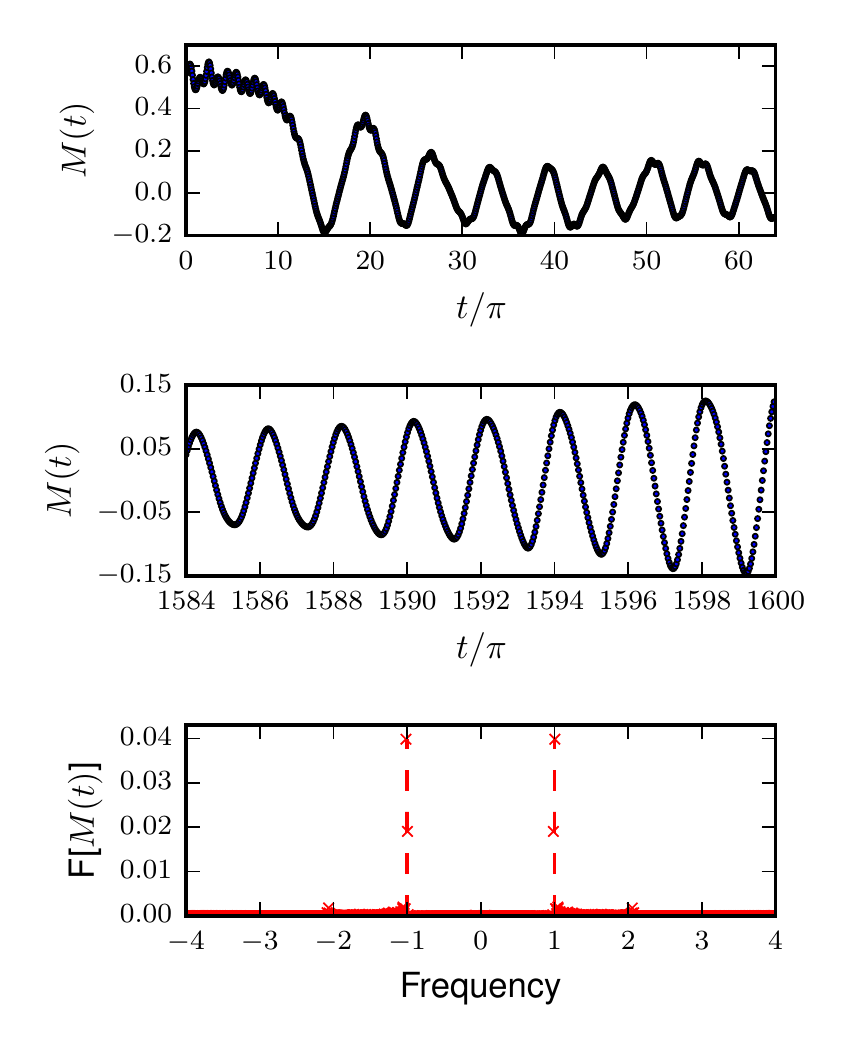}
\caption{ Top and middle panels: $M(t)$ vs number of periods $t/\pi$ at early and late times when the ordered ground state is driven starting at $t=0$. Bottom panel: Fourier spectrum of the time series between $t/\pi=1200$ and $t/\pi=1600$. Parameters: $d=2$, $r_0=-0.4$, $u=1$, $\Lambda=1$, $g=0.1$ and system size $L=100$.}
\label{Fig:MagFerroON}
\end{center}
\end{figure}

\begin{figure}[tb]
\begin{center}
\includegraphics{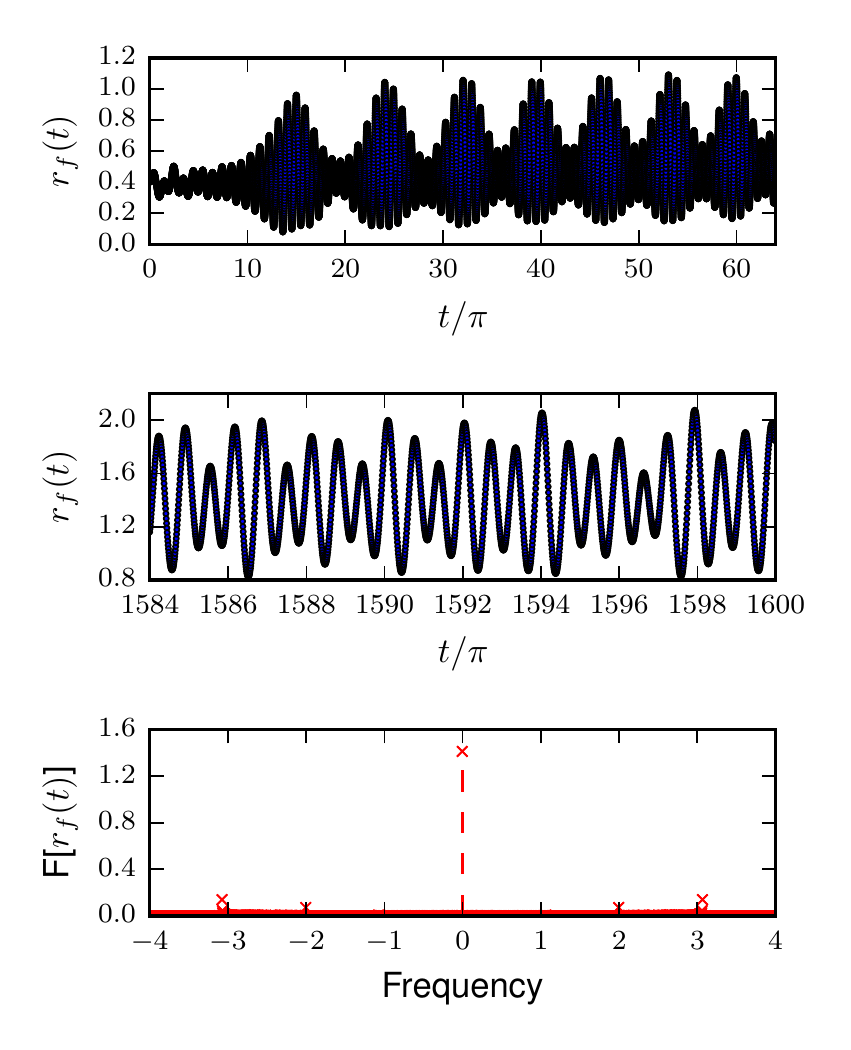}
\caption{ Top and middle panels: $r_f(t)$ vs number of periods $t/\pi$ at early and late times for the same parameters as Fig.~\ref{Fig:MagFerroON} is driven starting at $t=0$. Bottom panel: Fourier spectrum of the time series between $t/\pi=1200$ and $t/\pi=1600$. The energy density shows a similar behavior.}
\label{Fig:Phi2FerroON}
\end{center}
\end{figure}

\paragraph{Numerics---} 
Numerical simulations confirm the effective picture discussed above.
Consider the $d=2$ $O(N)$ model in its ferromagnetic ground state at $t=0$.
At time $t=0$, we turn on the drive.
The parameters are chosen such that the initial mode range includes the first band gap in the Mathieu spectrum.

Figs.~\ref{Fig:MagFerroON} and \ref{Fig:Phi2FerroON} show $M(t)$ and $r_f(t)$ in the time and frequency domain.
As expected, $M(t)$ decays to zero and $r_f(t)$ grows until $\bar{E}_0$ coincides with the bottom of the second band.
As the bottom of the second band corresponds to quasi-momentum $q=\pm1$, the period of $M(t)$ is twice the period of the drive (Fig.~\ref{Fig:MagFerroON} bottom panel).

The Fourier spectrum of $r_f(t)$ is also shown in the bottom panel of Fig.~\ref{Fig:Phi2FerroON}.
The dominant frequencies in the Fourier spectrum of $r_f(t)$ are seen to be at $\omega=0$ and $\omega=\pm 3$. 
The small weight at $\omega=\pm 3$ opens mini-gaps in the effective Mathieu spectrum at $q=\pm 1/2$, as expected for a Gaussian model driven at frequency one \footnote{The mode range must not include a mini-gap for stability.}.

\section{Discussion}
\label{Sec:Discussion}
Intuitively, a Floquet system stops absorbing energy from a monochromatic drive when a fraction of its modes saturate, as in hole burning.
This holds for fermions at the Gaussian level, as the fermionic modes do not interact, irrespective of whether the modes are spatially delocalized or localized.
For bosons however, individual modes can absorb energy indefinitely by parametric resonance at the Gaussian level.
The reader might expect that going beyond the Gaussian level in either case leads to indefinite heating, as interactions allow for the exchange of energy between modes.
One way to cut off this heating is with sufficient quenched disorder, so that the modes that are nearby in energy are far away in space and are unable to exchange energy and thermalize.
In this article, we discussed a different mechanism to cut off the heating, while still preserving enough symmetries to prevent full ergodicity.
The clean driven $O(N)$ model at infinite $N$ always has a finite energy density as $t\rightarrow \infty$, irrespective of the strength and fundamental frequency of the drive.

At finite $N$, on the other hand, the $O(N)$ model is believed to thermalize.
Previous work \cite{Berges:2007df} argues that corrections up to order $1/N^2$ are required to see true thermalization in sudden quenches.
This suggests that the driven $O(N)$ model could indefinitely heat once corrections to this order are included.
However, the time scale for this heating would be parametrically large in $N$, so that the steady state discussed in this article would be observable up to this time.
A detailed study of this question will be presented in a future work.

Recently, Citro et. al~\cite{Citro:2015aa} studied the thermalization of another scalar field theory: the driven ``many-body Kapitza pendulum'' or the driven sine-Gordon model in $d=1$.
They find that for parameters corresponding to both equilibrium phases---gapped and gapless---of the model, there is a critical frequency of the drive, below which the system heats indefinitely, but above which the heating stops. Their conclusions stem from various approximate
methods, of which one leads to equations that resemble the infinite $N$ equations studied in this paper. Currently we do not understand the difference between their results in the gapped/paramagnetic phase and our
results at low frequencies. We note that another of their methods---the application of the perturbative renormalization group to the Floquet Hamiltonian---is problematic as the stability of the Floquet Hamiltonian depends on the sign of the irrelevant terms. 
Understanding the precise connections between their results and ours will significantly clarify the steady state behavior of generic driven interacting bosonic systems.

\begin{acknowledgments}
The authors would like to thank D. Abanin, V. Khemani, A. Lazarides and R. Moessner for many Floquet discussions and C.R. Laumann and A. Polkovnikov for comments on a draft of this article. This work was supported by the National Science Foundation via Grant No. DMR-1311781 and Grant No. NSF PHY11-25915, the Alexander von Humboldt Foundation and the German Science Foundation (DFG) via the Gottfried Wilhelm Leibniz Prize Programme at MPI-PKS. Research at Perimeter Institute is supported by the Government of Canada through Industry Canada and by the Province of Ontario through the Ministry of Economic Development and Innovation.
\end{acknowledgments}

\bibliography{master}
\end{document}